# *Ab initio* phonon structure of *h*-YMnO$_3$ in low-symmetry ferroelectric phase.

**Short title:** Phonons of *h*-YMnO$_3$

Konstantin Z. Rushchanskii[*] and Marjana Ležaić

Peter Grünberg Institut, Forschungszentrum Jülich, D-52425 Jülich, Germany

We present a systematic first-principles study of the phonon spectrum of hexagonal YMnO$_3$ in ferroelectric (multiferroic) phase. We investigated in detail the low-energy phonon modes, their dispersion, symmetry, as well as the infrared optical properties of the crystal and determined the phonon density of states.

[*] email: k.rushchanskii@fz-juelich.de

INTRODUCTION

Hexagonal Yttrium manganese oxide (*h*-YMNO$_3$) is known from the 1960s to have ferroelectric (FE) and antiferromagnetic (AFM) transitions within a single phase [1]. Recent renewed interest in compounds combining several order parameters made it one of the model materials for investigation of the physics observed in multiferroics. In contrast to its orthorhombic phase, where ferroelectricity is induced by spin waves and is therefore very weak and observable only below $T = 30$ K [2], hexagonal phase is FE already below $T = 914$ K [3]. However, magnetic ordering occurs only at Néel temperature $T_N = 80$ K, at which a transition to non-collinear AFM state is observed [4].

Due to frustration of the magnetic spins localized on Mn atoms a lot of interesting physics was observed in these materials. Non-collinear AFM ground state has the same symmetry as the crystalline lattice [5]. Therefore, the linear magnetoelectric coupling is forbidden. However, experimentalists reported several interesting effects originating from the higher-order magnetoelectric coupling: optical second harmonic generation [6], reversal of AFM order parameter by switching the direction of ferroelectric polarization [6, 7], etc. Recently, exceptionally large influence of the magnetic ordering on atomic displacements and lattice parameters was observed, pointing to the giant magneto-elastic coupling in *h*-YMnO$_3$ [8]. Discovery of three magnon branches below T$_N$ using inelastic neutron scattering [9] opens new frontier for the investigation of spin-phonon interaction [10]. Neutron scattering experiments pointed out that short-range correlations between spins at Mn atoms exist also in the paramagnetic state [11]. This observation motivated a detailed investigation of the possible magnon-phonon interaction in the paramagnetic state of this material by means of far infrared and terahertz spectroscopy [12]. Additional absorption areas in low-energy regions were found, which could be associated with the multiphonon or phonon-paramagnon interaction. Hence, it is important to have information not only about the long-wavelength phonons, but also to know their dispersion in the whole Brillouin zone because of the short-range correlations of spins above Neel temperature, which is pronounced at the Brillouin zone edges.

To our knowledge, *ab initio* calculations of the full phonon spectra for *h*-YMnO$_3$ in low-symmetry ferroelectric phase is lacking in the literature. The large unit cell of the crystal, which contains 30 atoms and its low symmetry make such calculations computationally costly. Besides, strong on-site Coulomb interaction on magnetic Mn$^{3+}$ cations can affect the phonon energies and should be treated properly. In this paper we present the results and analysis of thorough calculations of the full phonon spectrum of *h*-YMnO$_3$ in ferroelectric phase.

CALCULATION DETAILS

Theoretical investigations of the structural and dynamical properties of *h*-YMnO$_3$ were performed within the spin-polarized local density approximation (LDA) [13]. We used projector-augmented potentials as implemented in Vienna *Ab initio* Simulation Package (VASP) [14-17]. The following electronic configurations of atoms were considered as valence: $4s^2 4p^6 5s^2 4d^1$ for Y, $3p^6 4s^2 3d^5$ for Mn and $2s^2 2p^4$ for O. The strong correlation effects on d electrons of Mn atoms were accounted by means of LDA+*U* approach [18, 19]. We used an on-site Coulomb parameter $U$ = 8.0 eV and Hund's exchange parameter $J_H$ = 0.88 eV as calculated in Ref. [20]. We didn't account for the spin-orbit interaction in our studies. To simulate paramagnetic spin alignment at Mn sites, we have used an A-type antiferromagnetic pattern, where spins are aligned ferromagnetically within the Mn honeycomb layers and in each layer along the hexagonal c-axis spins are flipped [21].

Plain-wave basis was determined by kinetic energy cutoff of 500 eV; Brillouin zone integration was performed by means of special k-points method with 4×4×2 Γ-centered *k*-point mesh. Structural parameters of unit cell were obtained by minimizing the Hellman-Feynman forces to a value smaller than 0.5 eV/Å.

Phonon calculations were performed by means of the force-constant method [22, 23]. Hellman-Feynman forces were calculated for atomic displacements of 0.04 Å in 2×2×2 supercell. For the supercell calculations we used a Γ-centered *k*-point mesh reduced to 2×2×1. The dynamical matrix for phonons at the arbitrary *q*-points situated along the high symmetry lines was constructed by a Fourier transform of the force constants, calculated at the Γ-point and for the Brillouin zone boundaries. Phonon mode wavenumbers and atomic displacement patterns for each *q*-point were obtained as eigenvalues and eigenvectors of the corresponding dynamical matrix.

THEORETICAL STRUCTURE

Crystalline structure of *h*-YMnO$_3$ in the ferroelectric phase is described by the space group P6$_3$cm [3]. The unit cell is built by the stacks of the MnO$_5$ trigonal bipyramid layers, separated along the *c*-axis by Y$^{3+}$ cations. MnO$_5$ bipyramid is formed by $Mn^{3+}$ cation situated in the center of the pyramid and surrounded by three planar oxygen atoms and two apical oxygens. Yttrium atoms, as well as planar oxygens, occupy two inequivalent Wyckoff positions, $a$ and $b$, whereas Mn cations and apical oxygens are in position $c$ (see Table 1). In-plane position $x$ of Mn is very close to 1 / 3, therefore Mn cations form nearly ideal triangular lattice, leading to the frustration of the magnetic spins.

Table 1. Theoretical and experimental [3] (presented in square brackets) relative atomic positions in $h$-YMnO$_3$. Hexagonal lattice parameters are $a$ = 6.0966 [6.1387] and $c$ = 11.41947 [11.4071] Angstroms, correspondingly.

| Atom | Wykoff position | x | y | z |
|---|---|---|---|---|
| Y1 | 2$a$ | 0 [0 | 0 0 | 0.52269 0.52172] |
| Y2 | 4$b$ | 2/3 [2/3 | 1/3 1/3 | 0.97873 0.98091] |
| Mn | 6$c$ | 0.33342 [0.3352 | 0 0 | 0.24749 0.24738] |
| O1 | 6$c$ | 0.30572 [0.3083 | 0 0 | 0.41156 0.4101] |
| O2 | 6$c$ | 0.63981 [0.6413 | 0 0 | 0.58346 0.5846] |
| O3 | 2$a$ | 0 [0 | 0 0 | 0.72339 0.7256] |
| O4 | 4$b$ | 2/3 [2/3 | 1/3 1/3 | 0.26815 0.2660] |

For the calculation of the phonon spectra all atomic positions of the structure have to be in their equilibrium, i.e. forces acting on the atoms have to be close to zero. We have preformed the relaxation of the crystallographic degrees of freedom of $h$-YMnO$_3$ and found that our theoretical values of crystallographic parameters are in a very good agreement with the available experimental data (see Table 1). The deviation of the calculated lattice parameters from the experimental one is -0.69 % for the in-plane periods and +0.11 % for the period along the hexagonal axis, which is within typical error of LDA approximation. Therefore, for the calculation of the phonons we used *ab initio* theoretical structure, obtained without additional constraints like, e.g., fixing the volume of the unit cell to its experimental value.

DYNAMICAL PROPERTIES

Calculated phonon spectrum of the ferroelectric $h$-YMnO$_3$ is presented in Figure 1. Clearly, the considered crystalline structure is dynamically stable over the entire Brillouin zone. The lowest-energy long-wavelength optical mode has the wavenumber of 96.5 cm$^{-1}$. The calculated acoustic modes have a dispersion typical for a rigid three-dimensional crystal, i.e. without features of membrane waves, which points to the strong interaction between MnO$_5$ bipyramides and Y cations. We have estimated the speed of sound for different directions: for the Γ-M direction a longitudinal wave has the speed of 7542 m/s, and transversal waves travel at 3443 and 2244 m/s. In the Γ-K direction the speed is 7370 m/s for the longitudinal, and 3442 and 2755 m/s for the transversal waves. Along the hexagonal axis the longitudinal wave has the speed of 8708 m/s, whereas the doubly degenerate transversal waves – 4237 m/s.

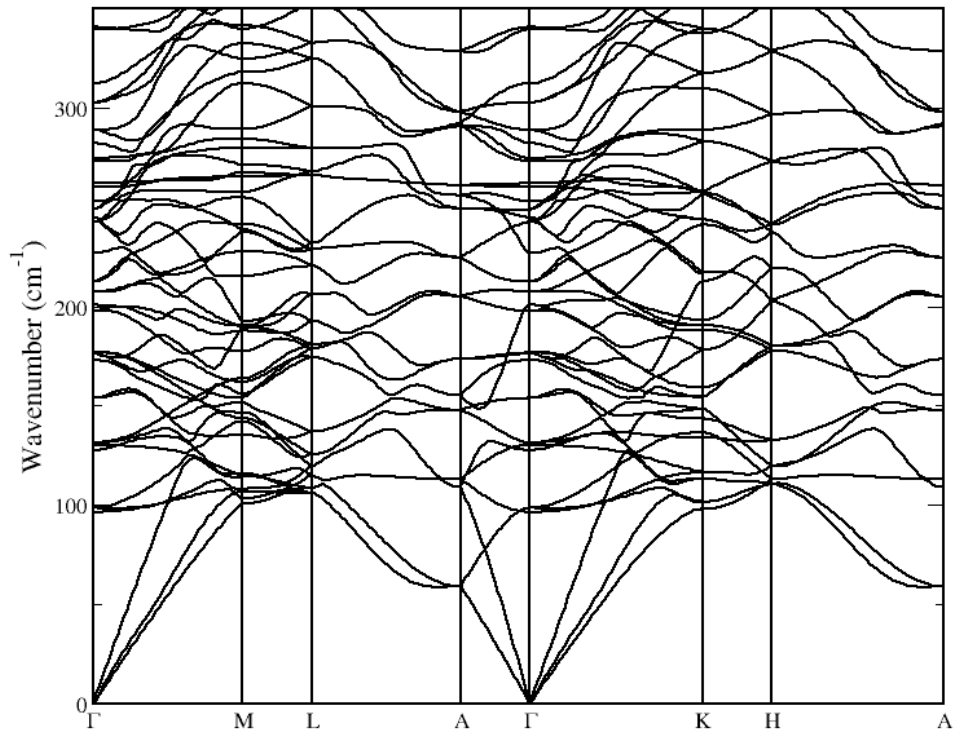

Figure 1. Phonon spectrum of $h$-YMnO$_3$ in the ferroelectric phase.

The experimental value of the energy of the magnon excitation has been found to be of 43 cm$^{-1}$ [10]. Due to an important role of the optical waves in the possible phonon-magnon interaction we have analyzed in detail the pattern of eigenvectors for the optical modes below 200 cm$^{-1}$. The results are collected in Table 2. Among the modes only E$_1$ and A$_1$ are polar (detailed group-symmetry analysis of phonons in Γ-point of $h$-YMnO$_3$ can be found in Ref. [24]). We have estimated the mode-plasma frequency values for the polar modes employing scalar

Born effective charges, calculated to be +3.6 |$e$| for Y, +3.2 |$e$| for Mn and -2.3 |$e$| for O [21]. Modes 153.9, 172.8 and 197.6 cm$^{-1}$ have significantly high values of mode-plasma frequencies, equal to 243, 117 and 291 cm$^{-1}$ respectively and therefore are dominating in the far infra-red spectra, whereas the mode at 176.6 cm$^{-1}$ is weak with mode-plasma frequency equal to 43 cm$^{-1}$. The obtained wavenumbers for these polar modes are in a very good agreement with those known from the low-temperature experiments [25]: 167 cm$^{-1}$ for $E_1$, 163 cm$^{-1}$ for $A_1$ and 197 cm$^{-1}$ for $E_1$, respectively. For the mode $E_1$ at the theoretical wavenumber 153.9 cm$^{-1}$, the displacements of Mn atoms have the largest impact, and they are out-of-phase to Y displacements. Mode $A_1$ with the theoretical wavenumber 172.9 cm$^{-1}$ is built mainly by out-of-phase displacements of Y atoms against the rigid MnO$_5$ bipyramides. Mode $E_1$ at 197.6 cm$^{-1}$ is built mostly by out-of-phase displacements of Mn atoms against the in-plane oxygens.

Table 2. Information on the low-energy optical phonons: wavenumber, symmetry and polarization of different atomic sublattices in the corresponding eigenvectors: "ab" means in-plane polarization, "c" means out-of-plane polarization, "ab/c" – mostly in-plane polarization with a small admixture of c-components, "c/ab" – mostly out-of-plane polarized displacement, 0 – sublattice does not contribute to the displacements.

| Mode | Wavenumbers, cm$^{-1}$ | Symmetry | Polarization of sublattices | | | | | | |
|---|---|---|---|---|---|---|---|---|---|
| | | | Y1 (2a) | Y2 (4b) | Mn (6c) | O1 (6c) | O2 (6c) | O3 (2a) | O4 (4b) |
| 1 | 96.5 | $A_2$ | 0 | c | ab | ab | ab | 0 | c |
| 2 | 98.6 | $E_2$ | ab | ab | ab/c | ab/c | ab/c | ab | ab |
| 3 | 127.7 | $B_2$ | c | c | ab/c | ab/c | ab/c | c | c |
| 4 | 130.4 | $B_1$ | 0 | c | ab | ab | ab | 0 | c |
| 5 | 131.4 | $E_2$ | ab | ab | ab/c | ab/c | ab/c | ab | ab |
| 6 | 153.9 | $E_1$ | ab | ab | ab/c | ab/c | ab/c | ab | ab |
| 7 | 172.9 | $A_1$ | c | c | ab/c | ab/c | ab/c | c | c |
| 8 | 176.2 | $E_2$ | ab | ab | ab/c | ab/c | ab/c | ab | ab |
| 9 | 176.7 | $E_1$ | ab | ab | ab/c | ab/c | ab/c | ab | ab |
| 10 | 197.6 | $E_1$ | ab | ab | c/ab | c/ab | c/ab | ab | ab |
| 11 | 200.7 | $B_2$ | c | c | c/ab | c/ab | 0 | c | c |

Low-energy phonons at the A-point (at the edge of the Brillouin zone) have the following wavenumbers: 59, 147, 173 and 205 cm$^{-1}$ (in-plane polarized modes) and 109, 113, and 155 cm$^{-1}$ (modes polarized along the hexagonal axis).

We have also calculated the total phonon density of states, as well as the Y, Mn and O sublattice contributions to it. As expected due to its largest mass, Y has the important impact only for the modes below 200 cm$^{-1}$, whereas the vibrations of the lightest oxygen dominate in the high-energy region above 350 cm$^{-1}$. High-energy excitations above 580 cm$^{-1}$ are separated by a small gap (about 30 cm$^{-1}$) from the rest of the spectrum.

Figure 2. Total and partial phonon density of states, calculated for ferroelectric *h*-

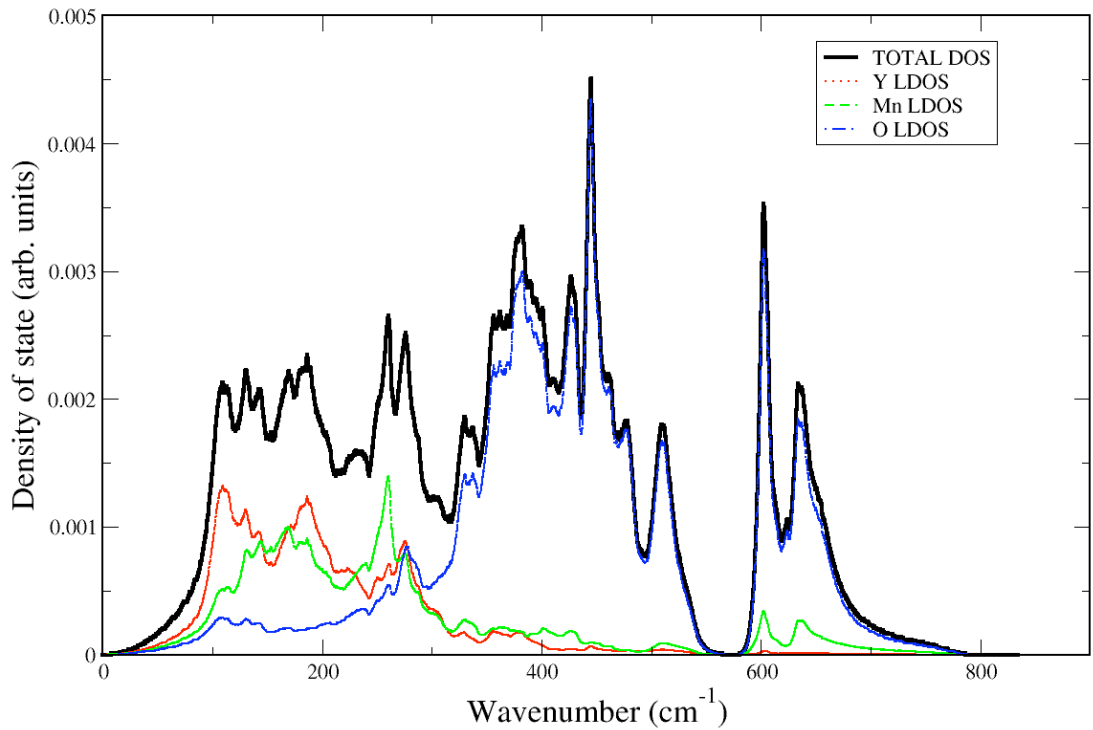

YMnO$_3$.

CONCLUSIONS

In this article we present the results of first-principles calculations of the phonon dispersion curves in the low-symmetry ferroelectric *h*-YMnO$_3$. We have found that the theoretically calculated crystalline structure of this material agrees well with the available experimental data. We have performed the analysis of the eigendisplacements pattern for the low-energy optic modes and calculated sound speed in different directions. Finally, we analyzed the phonon density of states and the contributions of Y, Mn and O to the vibrational spectra.

ACKNOWLEDGMENTS

For many fruitful discussions we thank Prof. Dr. Stanislav Kamba and Dr. Ivetta Slipukhina. We gratefully acknowledge the support by the Young Investigators Group Program of the Helmholtz association (Contract VH-NG-409), as well as the support of the Jülich Supercomputer Center.

REFERENCES

[1]     Bertaut EF, Mercier M: Structure magnetique de MnYO$_3$. Phys. Lett. 1963; 5: 27-29.
[2]     Fina I, Fabrega L, Marti X, Sanchez F, Foncuberta J: Magnetic switch of polarization in epitaxial orthorhombic YMnO$_3$ thin films. Appl. Phys. Lett. 2010; 97: 232905-3.
[3]     Abrahams SC: Ferroelectricity and structure in the YMnO$_3$ family. Acta Crystallogr. B. 2001; 57: 485-490.
[4]     Katsufuji T, Mori S, Masaki M, Moritomo Y, Yamamoto N, Takagi H: Dielectric and magnetic anomalies and spin frustration in hexagonal RMnO$_3$ (R=Y, Yb, and Lu). Phys. Rev. B 2001; 64: 104419-6.
[5]     Fiebig M, Frohlich D, Kohn K, Leute St, Lottermoser Th, Pavlov VV, Pisarev RV: Determination of the Magnetic Symmetry of Hexagonal Manganites by Second Harmonic Generation. Phys. Rev. Lett. 2000; 84: 5620–5623.
[6]     Fiebig M, Lottermoser Th, Frohlich D, Goltsev AV, Pisarev RV: Observation of coupled magnetic and electric domains. Nature 2002; 419: 818–820.
[7]     Choi T, Horibe Y, Yi TH, Choi YJ, Wu Weida, Cheong SW: Insulating interlocked ferroelectric and structural antiphase domain walls in multiferroic YMnO$_3$. Nature Mat. 2010; 9: 253-258.
[8]     Lee S, Pirogov A, Kang M, Jang KH, Yonemura M, Kamiyama T, Cheong SW, Gozzo F, Shin N, Kimura H, Noda Y, Park JG: Giant magneto-elastic coupling in multiferroic hexagonal manganites. Nature 2008; 451: 805-808.
[9]     Chatterji T, Ghosh S, Singh A, Regnault LP, Rheinstadter M: Spin dynamics of YMnO$_3$ studied via inelastic neutron scattering and the anisotropic Hubbard model. Phys. Rev. B. 2007; 76: 144406-10.
[10]    Petit S, Moussa F, Hennion M, Pailhès S, Pinsard-Gaudart L, Ivanov A: Spin Phonon Coupling in Hexagonal Multiferroic YMnO$_3$. Phys. Rev. Lett. 2007; 99: 266604-4.

[11] Demmel F, Chatterji T: Persistent spin waves above the Néel temperature in YMnO$_3$. Phys. Rev. B 2007; 76: 212402-4.

[12] Kadlec C, Goian V, Rushchanskii KZ, Kužel P, Ležaić M, Kohn K, Pisarev RV, Kamba S: Terahertz and infrared spectroscopic evidence of phonon-paramagnon coupling in hexagonal piezomagnetic YMnO$_3$. Phys. Rev. B 2011; 84: 174120-8.

[13] Perdew JP, Zunger A: Self-interaction correction to density-functional approximations for many-electron systems. Phys. Rev. B 1981; 23: 5048-5079.

[14] Kresse G, Hafner J: *Ab initio* molecular dynamics for liquid metals. Phys. Rev. B 1993; 47: 558-561.

[15] Kresse G, Furthmuller J: Efficient iterative schemes for *ab initio* total-energy calculations using a plane-wave basis set. Phys. Rev. B 1996; 54: 11169-11186.

[16] Blochl PE: Projector augmented-wave method. Phys. Rev. B 1994; 50: 17953-17979.

[17] Kresse G, Joubert D: From ultrasoft pseudopotentials to the projector augmented-wave method. Phys. Rev. B 1999; 59: 1758-1775

[18] Anisimov VI, Aryasetiawan F, Lichtenstein AI: First-principles calculations of the electronic structure and spectra of strongly correlated systems: the LDA+U method. J. Phys.: Condens. Matter 1997; 9: 767-808.

[19] Dudarev SL, Botton GA, Savrasov SY, Humphreys CJ, Sutton AP: Electron-energy-loss spectra and the structural stability of nickel oxide: An LSDA+U study. Phys. Rev. B 1998; 57: 1505-1509.

[20] Medvedeva JE, Anisimov VI, Korotin MA, Mryasov ON, Freeman AJ: The effect of Coulomb correlation and magnetic ordering on the electronic structure of two hexagonal phases of ferroelectromagnetic YMnO$_3$. J. Phys.: Condens. Matter 2000; 12: 4947-4958.

[21] van Aken BB, Palstra TTM, Filippetti A, Spaldin NA: The origin of ferroelectricity in magnetoelectric YMnO$_3$. Nat. Mater. 2004; 3: 164-170.

[22] Alfe D: PHON: A program to calculate phonons using the small displacement method. Comp. Phys. Commun. 2009; 180: 2622-2633.

[23] Kunc K, Martin RM: *Ab Initio* Force Constants of GaAs: A New Approach to Calculation of Phonons and Dielectric Properties. Phys. Rev. Lett. 1982; 48: 406-409.

[24] Iliev MN, Lee HG, Popov VN, Abrashev MV, Hamed A, Meng RL, Chu CW: Raman- and infrared-active phonons in hexagonal YMnO$_3$: Experiment and lattice-dynamical calculations. Phys. Rev. B 1997; 56: 2488–2494.

[25] Zaghrioui M, Ta Phuoc V, Souza RA, Gervais M: Polarized reflectivity and lattice dynamics calculation of multiferroic YMnO$_3$. Phys. Rev. B 2008; 78: 184305-6.